# Quantum Conductance Oscillations in Metal/Molecule/Metal Switches at Room Temperature

Feng Miao[1], Douglas Ohlberg[2], Duncan Stewart[2], R. Stanley Williams[2] and Chun Ning Lau[1*]

1. Department of Physics and Astronomy, University of California, Riverside, CA 92521

2. Quantum Science Research, Hewlett-Packard Laboratories, Palo Alto, CA 94304

* lau@physics.ucr.edu

**Abstract**

Conductance switching has been reported in many molecular junction devices, but in most cases has not been convincingly explained. We investigate conductance switching in Pt/stearic acid monolayer/Ti devices using pressure-modulated conductance microscopy. For devices with conductance $G \gg G_Q$ or $G \ll G_Q$, where $G_Q = 2e^2/h$ is the conductance quantum, pressure-induced conductance peaks <30 nm in diameter are observed, indicating the formation of nanoscale conducting pathways between the electrodes. For devices with $G \sim$ 1- 2 $G_Q$, in addition to conductance peaks we also observed conductance dips and oscillations in response to localized pressure. These results can be modeled by considering interfering electron waves along a quantum conductance channel between two partially transmitting electrode surfaces. Our findings underscore the possible use of these devices as atomic-scale switches.

Nanoscale switches are extremely important for both memory and logic applications in all future ultra-small integrated circuits. The vast majority of nanoscale switches reported are based on storage and manipulation of electrical charge. However, as conventional devices approach the size of molecules and atoms, the probability of undesired charge tunneling or charge leakage increases exponentially, rendering reliable confinement and control of charges ever more challenging. In contrast, devices that are based on the movement of atoms and molecules present a solution to this challenge; they can be engineered at the atomic level to control and maintain switch states. Such devices include nanomechanical switches [1--5] and nanoscale "atomic" switches incorporating ionic conductors Ag and $Ag_2S$ [6].

Another candidate for these atomic switches are metal/molecular monolayer/metal heterostructures with conductances tunable by applied voltage or current. Our previous work demonstrated that conductance switching in Pt/stearic acid monolayer/Ti devices arises from the formation and dissolution of nanoscale conductance channels within the junction, likely due to electro-chemical reaction of the Pt and Ti electrodes [7, 8]. This is in contrast to a number of molecule-specific switching mechanisms found in molecular devices, such as redox-induced configuration change of molecules [9--12], fluctuation in the bond between metal and molecules [13], or charge transfer [14, 15]. The Pt/C18/Ti devices are thus nanoscale switches based on storage and manipulation of atoms, wherein the molecular layer acts as a porous medium through which atoms or ions can diffuse back and forth.

Here we report the reversible formation of quantum conductance channels in these devices, and evidence of quantum coherence of electrons at room temperature. Using the scanned probe technique of pressure-modulated conductance microscopy [7], we observe individual nanoscale conductance dips and oscillations in response to local mechanical pressure applied by an AFM tip. Such conductance dips are only observed in devices with conductance between 1 and 2 conductance quantum $G_Q$, where $G_Q=2e^2/h$ or ≈80μS. These results can be satisfactorily modeled by resonant electron transmission across quantum conductance channels that partially or completely bridge the top and bottom electrodes; these channels may be quantum point contacts or alignment of oxide deficiency sites. Our results underscore the small size of the active area responsible for switching, suggesting that the devices can be dramatically shrunk in size for nanoscale memory and logic applications.

The molecular devices consist of a Langmuir-Blodgett monolayer of stearic acid ($C_{18}H_{36}OH$) molecules sandwiched between top and bottom metal electrodes. The monolayer is 2.6± 0.2 nm in thickness, as determined by ellipsometry. The bottom electrodes are 300 nm of platinum, and the top electrodes are 10 - 20 nm of either titanium or chromium, followed by 5 nm of platinum. The devices can be switched on and off by applying appropriate voltages (Fig. 1b). The switching mechanism was investigated using the pressure-modulated conductance microscopy(PCM) [7, 8], in which an AFM tip is used to locally apply mechanical pressure while the conductance of the device is monitored. Plotting the device conductance as a function of the tip position yields an image of the electrical response of the device induced by local pressure (Fig. 1a). All PCM measurements are performed at room temperature. As reported previously [7, 8, 16], when the devices are in a high conductance state (R<~100kΩ), we observed individual nanoscale conductance peaks, or "switching centers", in response to mechanical pressure applied by an AFM tip (Fig. 1b, upper insets); when the same devices are switched "off", these peaks disappear (Fig. 1b, lower insets). These switching centers are the dominant conductance pathways, and may be most simply modeled as nanoscale asperities that partially or completely bridge the electrodes. The perfect correlation between the appearance

(disappearance) of a switching center and the switching "on" ("off") of a device indicates that conductance switching arises from the formation and dissolution of such nanoscale pathways.

To further investigate these switching centers, we performed PCM in more than 40 devices, and systematically examined the amplitude of the device's mechano-electrical response at a fixed compressive strain. The results are summarized in Fig. 2, which plots $\Delta g$, the relative change in conductance induced by pressure, as a function of the unperturbed conductance $G(0)$. Here $\Delta g = \frac{G(\varepsilon_{zz}) - G(0)}{G(0)}$, where $G(\varepsilon_{zz})$ is the conductance of the device under compressive strain $\varepsilon_{zz}$. To estimate $\varepsilon_{zz}$ experienced by the monolayer and any nano-asperities, we model the effect of the AFM tip using classical elasticity theory [19]. For a point force $F$ at the origin applied to a semi-infinite elastic medium, the local compressive strain $\varepsilon_{zz}$ at a depth $d$ inside the medium is approximately

$$\varepsilon_{zz}(x,y) \sim \frac{3}{2\pi E} \frac{d^3}{\left(x^2 + y^2 + d^2\right)^{5/2}} F \qquad (1)$$

Here E is the Young's modulus of the medium, ~100GPa for both the metals and the monolayer [20--23], $(x, y)$ are the co-ordinates of the point under consideration, and $d$ is taken to be the thickness of the top electrode. For $F=0.5\mu N$ and $d=15$ nm, we estimate that the monolayer is compressed by 1%. For the data shown in Fig. 2, depending on the thickness of top electrodes, the applied force ranges from 0.5 to 1.5 $\mu N$, yielding $\varepsilon_{zz}(0,0) \sim 1\%$.

The data points in Fig. 2 clearly fall into three regimes. For very resistive devices with $G << G_Q$, $\Delta g$ is relatively large, ~10-30%, whereas very conductive devices ($G>2G_Q$) has $\Delta g$ <~3%. (we will consider devices with intermediate conductance later). Both behaviors can be quantitatively understood by considering the effect of local pressure exerted on the nano-asperities. For very resistive devices, we note that generally, for electrical conduction across an insulator with thickness $L$, the conductance $G \propto e^{-\beta L}$, where $\beta$ is a parameter that may depend on the barrier height and temperature. Upon applied pressure, $L$ will be decreased from the initial value $L_0$ by an amount

$$\delta L = L - L_0 = \varepsilon_{zz} L_0 \qquad (2)$$

Thus, for devices with $G << G_Q$, the nano-asperity does not completely bridge the gap, $\Delta g = \frac{G(L_0 - \delta L)}{G(L_0)} - 1 \approx \beta \cdot \delta L$. Using $L_0 \sim 2$ nm and $\beta \approx 1/\text{Å}$ for alkane molecules [20, 24], we estimate that $\Delta g \sim 25\%$, as measured experimentally. On the other hand, for devices with $G >> G_Q$, there are many transmitting channels. Hence electrical response under compression is small, arising mainly from increasing the cross-sectional area. As the detailed configurations of the conducting pathways are not known, we can estimate the magnitude of $\Delta g$ by modeling the conducting pathway as either (1) a ballistic point contact with conductance $G \approx \frac{2e^2}{h} \frac{A}{\lambda_F^2}$ [25, 26], or (2) a diffusive contact with $G = \sigma \frac{A}{L}$. Here $\lambda_F$ is the Fermi wave length, and $A$ and $L$ are the total cross sectional area and length of the pathway, respectively. Assuming a Poisson's ratio of 0.3, both models yield $\Delta g \sim 1\%$, in agreement with the experimental data.

Now we focus exclusively on devices with an unperturbed conductance between 80 μS and 160 μS, or 1-2 $G_Q$. Two features immediately distinguish this regime from the other devices: (1) a large spread in values of $\Delta g$, ranging from a few percent to more than 30%; (2) in addition to conductance peaks, we also observed negative values of $\Delta g$, that is, nanoscale *dips* in conductance in response to local pressure (Fig. 3b). The latter feature is especially counter-intuitive, as decreasing the inter-electrode distance leads to a *decrease* in current.

Such inverse switching centers, or conductance dips with compression, are difficult to explain in terms of transport across either a tunnel barrier (i.e. with very small transmission coefficient), or across a number of high transmission channels. The fact that inverse switching centers are found only in devices with G~1-2 $G_Q$ is particularly striking. It is reminiscent of the conductance parity oscillation observed in the final conductance plateau in metal point contacts formed by mechanically controlled break junctions (MBJ) [26--28]. *Our data thus suggest the formation of one or two highly-transmitting quantum conductance channels between the top and bottom electrodes.* These channels may be a linear chain of atoms (i.e. a metal point contact), or perhaps a column of aligned oxygen deficiency sites in titanium oxide[17, 18, 29]. Upon applied pressure, the atoms undergo slight re-arrangement, altering the amount of inter-atomic orbital overlap and thus reducing the transmission coefficient of the conductance channel(s). Such conformation-induced change in orbital overlap may take place via a number of atomic configurations. Here we adopt a particularly straightforward model to aid our quantitative understanding of the data: we assume the conducting channels bridging between the electrodes form a small Fabry-Perot cavity for electrons. As electron waves propagate through the atomic chain between two partially reflecting barriers, the incoming and multiply-reflected electrons paths interfere and give rise to periodic oscillations in the transmission coefficient as a function of inter-electrode spacing.

Quantitatively, by matching the boundary conditions of a 1D wave function of a particle propagating through two barriers, the transmission coefficient is given by [28]

$$T = \frac{16\gamma^2}{(1+\gamma)^4 + (1-\gamma)^4 - 2(1+\gamma)^2(1-\gamma)^2 \cos(2k_2 L)} \qquad (3)$$

where $\gamma = k_2/k_1$, $L$ is the length of the channel, and $k_2$ and $k_1$ are the wave vectors in the electrodes and in the atomic chain, respectively. Here we treat $k_2$ and $L$ as independent variables. Also, for simplicity we assume that the electrons in the two electrodes have the same wave vector (different wave vectors will modulate the absolute values of *T*). The resulting graph is a periodic function of *L*, as plotted in Fig. 3a, using reasonable parameters for metals, $k_2=10^{10}$ m$^{-1}$, $\gamma=1$ (solid lines) and $\gamma=0.5$ (dotted lines), respectively. Thus, as the compressive pressure from the AFM tip diminishes *L*, or the effective size of the cavity, the total transmission can either increase or decrease, depending on $L_0$, the initial value of *L*. In fact, it is easily inferred from the plot that, assuming a random distribution of $L_0$, *T* is equally likely to be enhanced or reduced by slight modulation in *L*. We therefore expect roughly same number of conductance peaks and dips for devices in this regime. This is indeed borne out by the experimental data: for a total of 19 devices with *G(0)* between 1 and 2 $G_Q$, we observed 9 normal and 10 inverse switching centers.

Moreover, we can reproduce the conductance dips in the PCM images by substituting Eq. (1) and (2) into (3) and obtaining a transmission coefficient as a function of tip position *(x, y)*. Assuming 1-channel transmission and taking into account electron spin, we have

$$G(x,y)=2e^2/h\ T(x,y) \qquad (4)$$

Assuming the initial length is $L_0=1.2$nm, the resulting conductance (4) is plotted as a function of tip position *(x,y)* (Fig. 3c). The resemblance between the data and the simulation is satisfactory. Thus our simple model can adequately explain observation of the inverse switching centers. Even though it is likely not a unique model and our choice of parameters is somewhat heuristic – in fact, any mechanism that gives rise to peaks and valleys in the transmission coefficient will yield similar results – it effectively explains a number of features observed in the data.

A crucial test of this model occurs in the case that the initial length $L_0$ only slightly exceeds the value corresponding to a local transmission minimum or maximum. We then expect to observe conductance *oscillations* with increasing pressure: as the AFM tip approaches the conducting channel, *L* decreases and reaches a minimum at the center, while *T(L)* and hence *G* pass through the local extremum and then reverse slope. Such conductance oscillations will appear as rings in the PCM image. Indeed, these rings are observed in a four of the devices with conductance $G \sim 1\text{-}2G_Q$ (but not in those with G>>GQ or G<<GQ). Both "M"-shaped (i.e. with a center peak, Fig. 4a, left panel) and "W"-shaped (i.e. with a center dip, Fig. 4b, right panel) ring are observed, corresponding to a local maximum and minimum in *T(L)*, respectively. We can also reproduce these features with our simple model, by using $L_0$=1.45nm and $L_0$=1.59nm, respectively (Fig. 4a and b, right panels).

A further key prediction of this model is the ability to manipulate the shape of the switching center by modulating the applied pressure. For one device with $G_Q < G < 2G_Q$, we succeeded in this shape manipulation. At a maximum stress $\varepsilon_{zz}$ of 2.4%, the device shown in Fig. 3b displayed an inverse switching center, with $\Delta g \sim -12\%$ (cross-section in Fig 4c top). Remarkably, when imaged at an increased stress of 6.5%, the exact same switching center appeared with oscillations as a "W"-shaped ring (Fig. 4c).

It should now be clear why conductance oscillations are not observed in devices with $G<<G_Q$ or $G>>G_Q$. In the former case, the transport is dominated by the exponential dependence of the inter-electrode spacing; any oscillation arising from electron interference between the electrodes will be miniscule compared to the exponential increase in current. For $G>>G_Q$, there are many highly transmitting channels. The effect of tuning the cavity length is likely to be different for different channels, and therefore will largely average to zero, hence again rendering the conductance dominated by the increase from larger cross-sectional areas.

In conclusion, pressure-modulated conductance microscopy has revealed conductance peaks, dips and oscillations in metal/molecular monolayer/metal junction devices. These phenomena can be readily explained by the interfering electron waves within quantum conductance channels in the device. This quantum nature of the devices underscores that the active switching area is atomic in scale, and suggests that this class of junction device can be shrunk almost to the atomic limit for future ultra-scaled memory and logic applications.

**Reference**

1. H. G. Craighead, Science **290**, 1532 (2000).

2. V. Sazonova, Y. Yaish, H. Ustunel, et al., Nature **431**, 284 (2004).

3. M. D. LaHaye, O. Buu, B. Camarota, et al., Science **304**, 74 (2004).

4. Q. S. Zheng and Q. Jiang, Phys. Rev. Lett. **88** (2002).

5. Y. T. Yang, K. L. Ekinci, X. M. H. Huang, et al., Appl. Phys. Lett. **78**, 162 (2001).

6. K. Terabe, T. Hasegawa, T. Nakayama, et al., Nature **433**, 47 (2005).

7. C. N. Lau, D. R. Stewart, R. S. Williams, et al., Nano Lett. **4**, 569 (2004).

8. C. N. Lau, D. R. Stewart, M. Bockrath, et al., Appl. Phys. A, **80**, 1373 (2005).

9. C. P. Collier, J. O. Jeppesen, Y. Luo, et al., J. Am. Chem. Soc. **123**, 12632 (2001).

10. C. P. Collier, G. Mattersteig, E. W. Wong, et al., Science **289**, 1172 (2000).

11. C. P. Collier, E. W. Wong, M. Belohradsky, et al., Science **285**, 391 (1999).

12. Z. J. Donhauser, B. A. Mantooth, K. F. Kelly, et al., Science **292**, 2303 (2001).

13. G. K. Ramachandran, T. J. Hopson, A. M. Rawlett, et al., Science **300**, 1413 (2003).

14. J. Chen, W. Wang, M. A. Reed, et al., Appl. Phys. Lett. **77**, 1224 (2000).

15. J. M. Seminario, A. G. Zacarias, and J. M. Tour, J. Am. Chem. Soc. **122**, 3015 (2000).

16. D. R. Stewart, D. A. A. Ohlberg, P. A. Beck, et al., Nano Lett. **4**, 133 (2004).

17. E. M. Lifshitz and L. D. Landau, *Theory of Elasticity* (Butterworth-Heinemann.

18. D. J. Wold and C. D. Frisbie, J. American Chem. Soc. **123**, 5549 (2001).

19. Y. S. Leng and S. Y. Jiang, J. Chem. Phys. **113**, 8800 (2000).

20. A. B. Tutein, S. J. Stuart, and J. A. Harrison, J. Phys. Chem. B **103**, 11357 (1999).

21. D. L. Patrick, J. F. Flanagan, P. Kohl, et al., J. Am. Chem. Soc. **125**, 6762 (2003).

22. D. J. Wold and C. D. Frisbie, J. Am. Chem. Soc. **122**, 2970 (2000).

23. A. M. Bratkovsky and S. N. Rashkeev, Phys. Rev. B **53**, 13074 (1996).

24. N. Agrait, A. L. Yeyati, and J. M. van Ruitenbeek, Phys. Rep. **377**, 81 (2003).

25. J. M. Krans, C. J. Muller, I. K. Yanson, et al., Phys. Rev. B **48**, 14721 (1993).

26. E. Yagi, R. R. Hasiguti, and M. Aono, Phys. Rev. B **54**, 7945 (1996).

27. M. Aono and R. R. Hasiguti, Phys. Rev. B **48**, 12406 (1993).

28. R. H. M. Smit, C. Untiedt, G. Rubio-Bollinger, et al., Phys. Rev. Lett. **91** (2003).

29. J.J. Blackstock, W.F. Stickle, C.L. Donley, et al., *J. Phys. Chem. B* **111**, 16 (2006).

FIG. 1. (a) Experimental Setup of pressure-modulated conductance microscopy. The conductance of the device is monitored while an AFM tip applies local mechanical pressure. (b). Switching characteristics of the device (arrows indicating hysteresis direction). Upper inset: a switching center appears in the PCM image when the device is switched into the "on" state. Lower inset: the switching center disappears when the device is switched to the "off" state. Scan Size: 10x2.5μm.

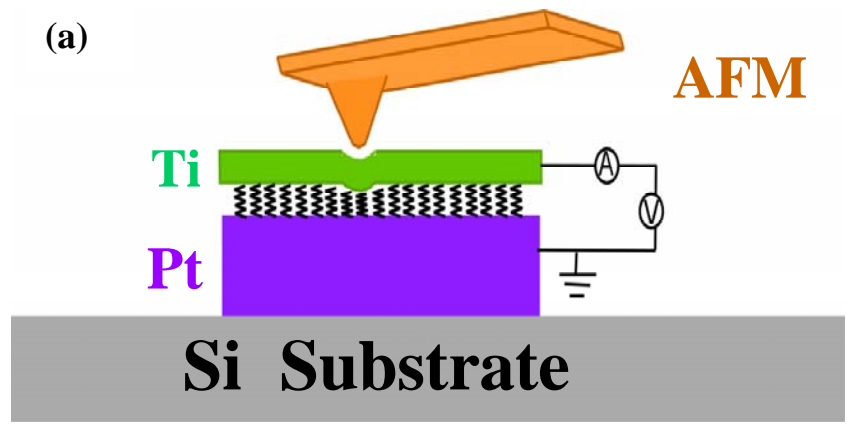

**(b)**

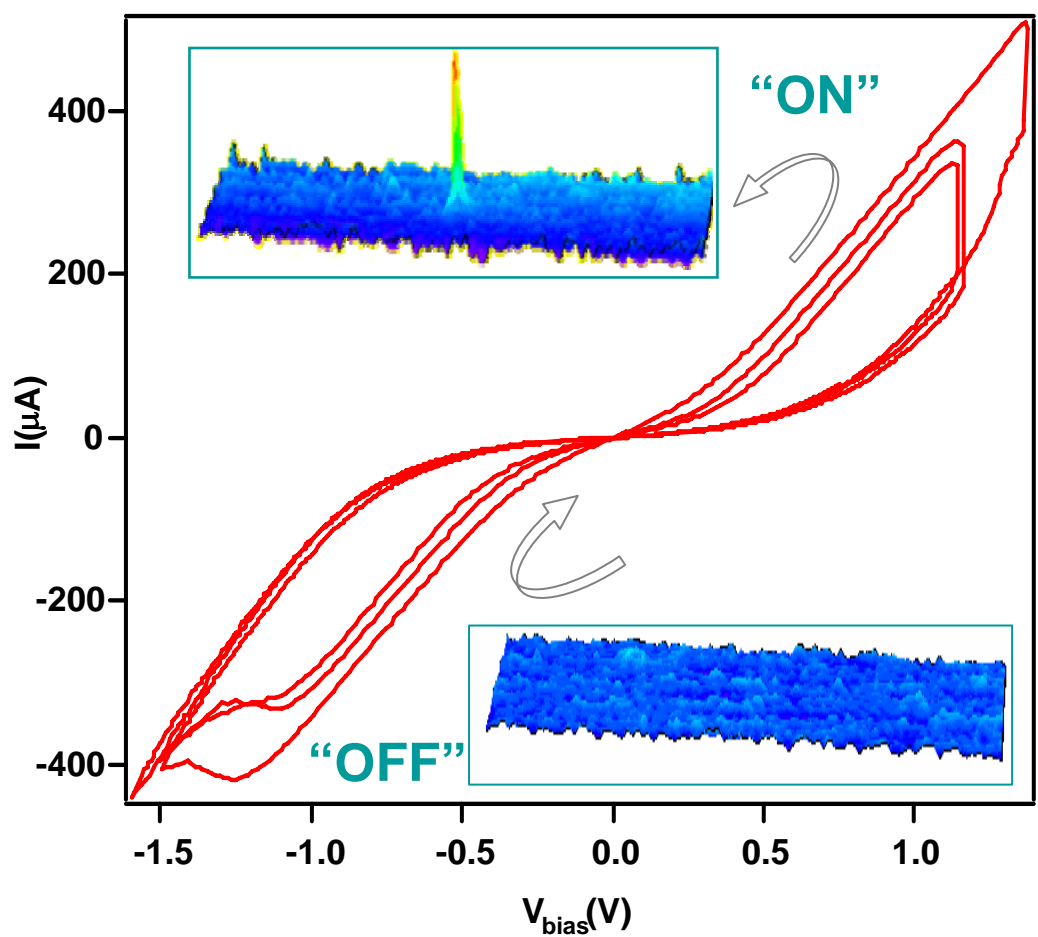

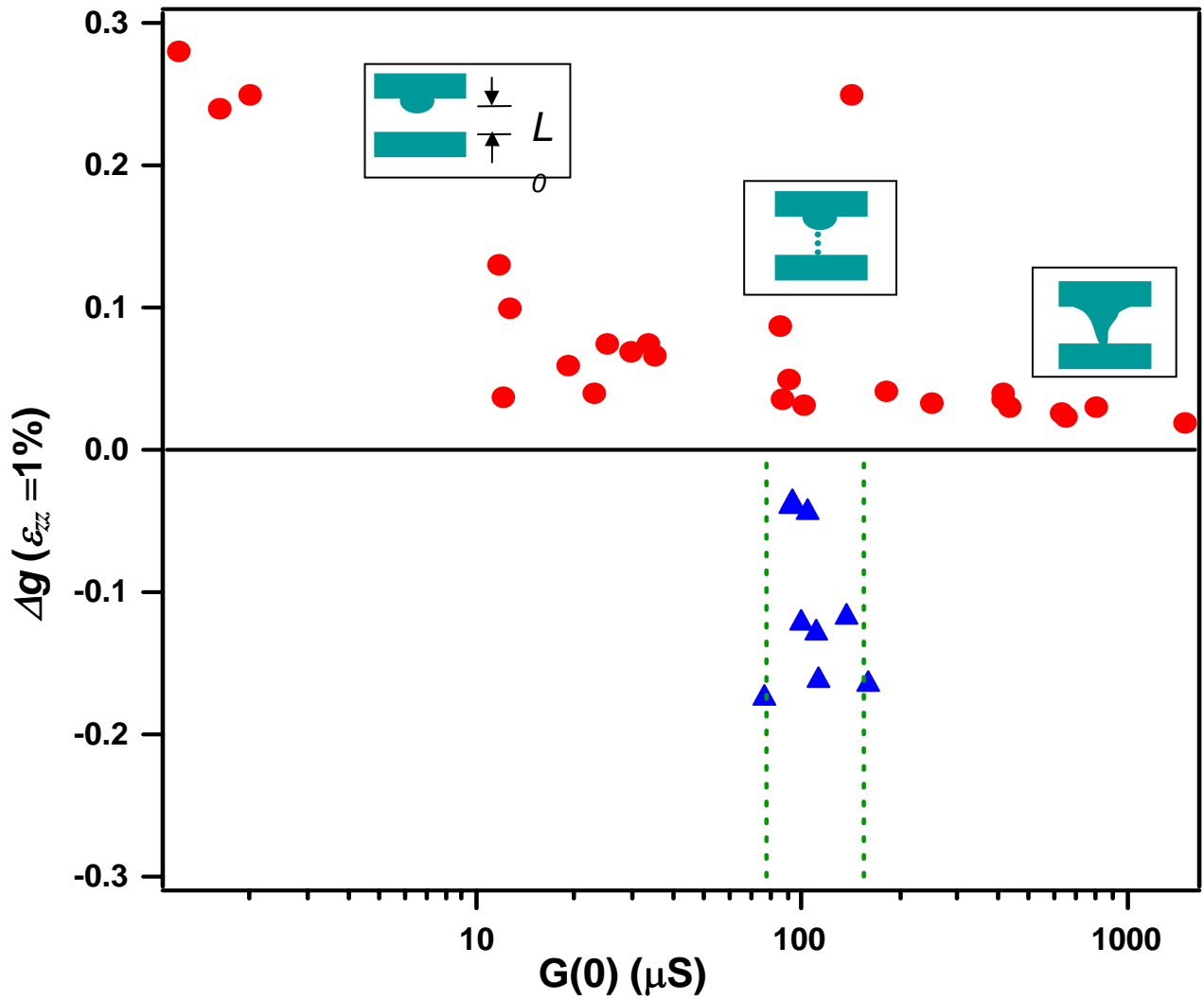

FIG. 2. Relative change in conductance under 1% compressive strain as a function of the unperturbed conductance $G(0)$. The dotted lines denote $G = G_Q$ and $G = 2 G_Q$, respectively. Insets: schematic representation of microscopic atomic configurations for devices with $G \ll G_Q$, $G \sim G_Q$, and $G(0) \gg G_Q$ respectively.

FIG. 3. (a). Periodic oscillations in the transmission coefficient as a function if inter-electrode distance, calculated using Eq. (4) and $k_2=10^{10}$ m$^{-1}$. Solid line: $\gamma=1$. Dotted line: $\gamma=0.5$. (b) Experimental data (left) and model (right) showing the conductance dip as a function of the tip position, with $\varepsilon_{zz}(0,0)=3.8\%$. Color scale: $G/G(0)$. The simulation is calculated using Equ. (4), $\gamma=0.5$ and $L_0=1.15$ nm.

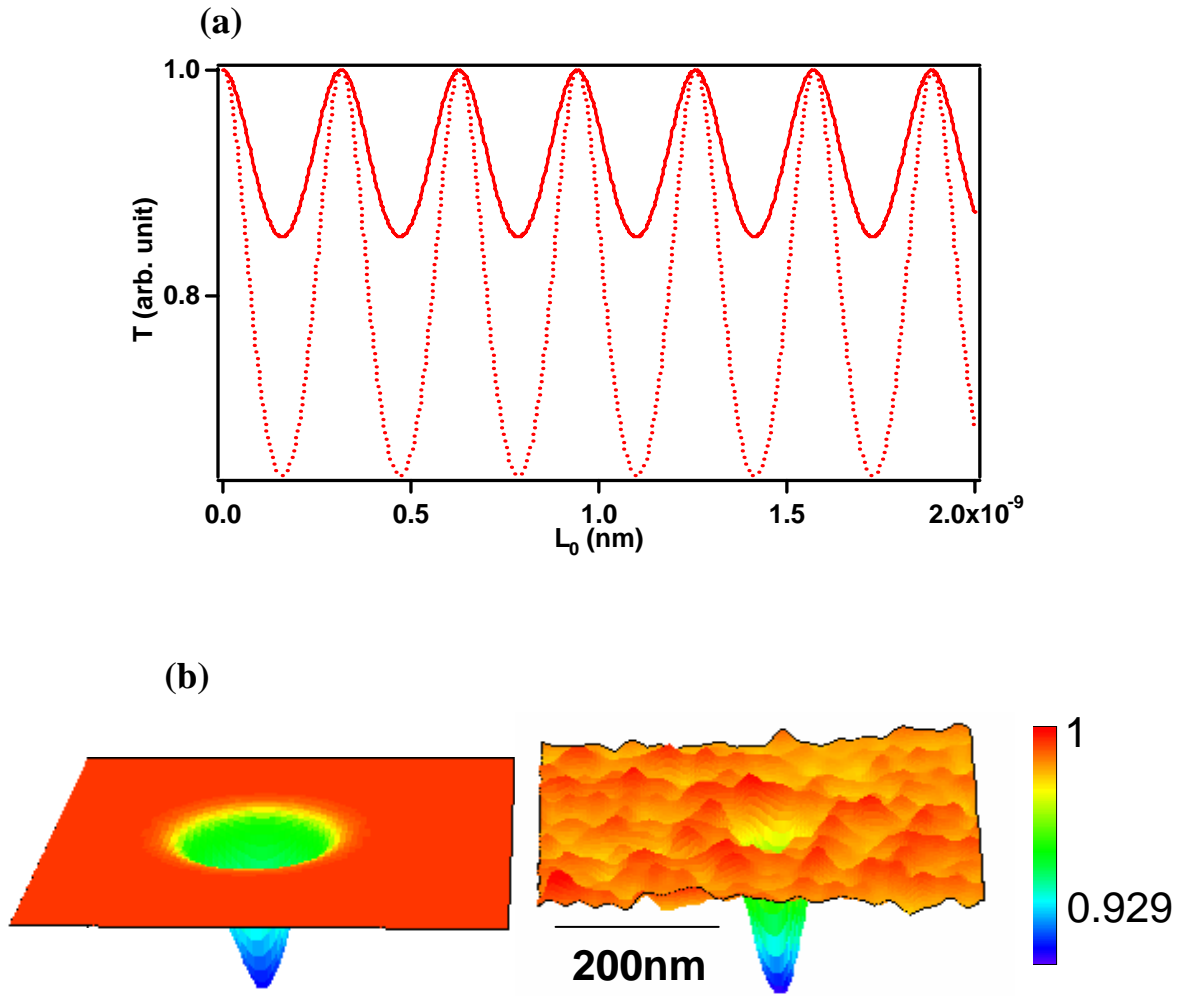

FIG. 4. (a) PCM image exhibiting W-shaped conductance oscillation as a function of tip position. Color and vertical scale: $G/G(0)$. Left Panels: data taken at $\varepsilon_{zz}(0,0) = 6\%$. Right Panels: Simulation calculated using Eq. (4), $\gamma=0.5$ and $L_0=1.45$. (b) M-shaped conductance oscillation observed in another device. The data are taken at $\varepsilon_{zz}(0,0) = 2\%$, and simulation calculated using $\gamma=0.5$ and $L_0=1.59$. (c). Line traces through experimental data from a switching center at different compressive strains: $\varepsilon_{zz}(0,0) = 2.4\%$ (upper) and 6.5% (lower trace).

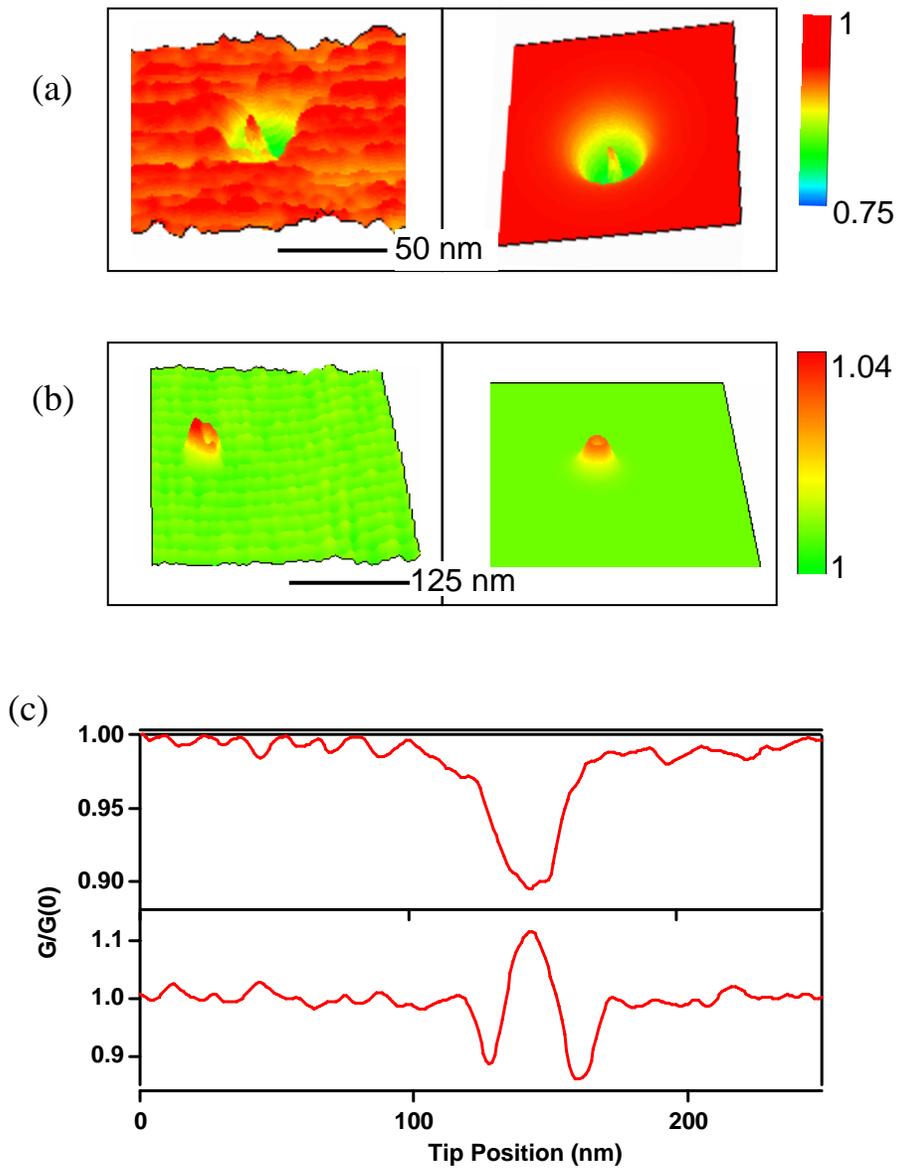